# How do referees integrate evaluation criteria into their overall judgment? Evidence from grant peer review

Sven E. Hug, University of Zurich, sven.hug@uzh.ch

Abstract: Little is known whether peer reviewers use the same evaluation criteria and how they integrate the criteria into their overall judgment. This study therefore proposed two assessment styles based on theoretical perspectives and normative positions. According to the case-by-case style, referees use many and different criteria, weight criteria on a case-by-case basis, and integrate criteria in a complex, non-mechanical way into their overall judgment. According to the uniform style, referees use a small fraction of the available criteria, apply the same criteria, weight the criteria in the same way, and integrate the criteria based on simple rules (i.e., fast-and-frugal heuristics). These two styles were examined using a unique dataset from a career funding scheme that contained a comparatively large number of evaluation criteria. A heuristic (fast-and-frugal trees) and a complex procedure (logistic regression) were employed to describe how referees integrate the criteria into their overall judgment. The logistic regression predicted the referees' overall assessment with high accuracy and slightly more accurately than the fast-and-frugal trees. Overall, the results of this study support the uniform style but also indicate that the uniform style needs to be revised as follows: referees use many criteria and integrate the criteria using complex rules. However, and most importantly, the revised style could describe most – but not all – of the referees' judgments. Future studies should therefore examine how referees' judgments can be characterized in those cases where the uniform style failed. Moreover, the evaluation process of referees should be studied in more empirical and theoretical detail.

Keywords: academic peer review, peer review theory, fast and frugal heuristics, commensuration bias, evaluation criteria, grant funding

## 1. Introduction

In academia, the attribution of value and worth as well as the allocation of reward usually involves peer review processes (Hamann & Beljean, 2017; Zuckerman & Merton, 1971). Peer review is thus seen as central to the practice of research (Baldwin, 2020; Johnson & Hermanowicz, 2017) and pervades modern scholarship.[1] For example, peer review is used in grant funding, scholarly communication (journal articles, books, conference contributions), preregistration of studies, hiring and promotion processes, institutional evaluations, or in the conferral of awards and honors. Given the ubiquity and centrality of the process, it is not surprising that a large literature on peer review has been published (Batagelj et al., 2017; Grimaldo et al., 2018). But despite more than 50 years of research, our theoretical understanding of peer review remains very limited (Bornmann, 2008; Chubin & Hackett, 1990; Gläser & Laudel, 2005; Hirschauer, 2004; Hug, 2022; Reinhart & Schendzielorz, 2021). For instance, we have made little progress with respect to the following broad questions: Why and how has peer review evolved? What are the causal mechanisms that produce peer review phenomena, most notably the many supposed or confirmed biases? How are peer review and the broader contexts in which it is embedded interrelated (organizations, scientific communities, science, society, economy, government)? Why and how does peer review work, and what are its effects? Which interventions and innovations improved peer review, how did they work, and for whom?[2] Due to this theoretical deficit, Hug (2022) encouraged peer review researchers to be more theoretically engaged and to theorize the many aspects of peer review.

The current study focuses on the broad question of how peer review works and, more specifically, on the evaluation process employed by reviewers/referees.[3] With respect to this process, little is known about whether referees use the same evaluation criteria (Arvan et al., 2022; Hug & Ochsner, 2022). Moreover, recent research has raised the question of how referees convert scores along heterogenous evaluation criteria into an overall scale of value (Erosheva et al., 2020; Heesen, 2019; Lee, 2015), or more generally, how referees integrate evaluation criteria into their overall judgment. In this study, I therefore focused on the following research questions: Do referees use the same criteria in their evaluation process? How do referees integrate the criteria into their overall judgment? Specifically, I linked these two questions with theoretical perspectives and normative positions, that is, the three-phase model of editorial judgment (section 1.1), the ideal of impartiality (section 1.2), commensuration bias (section 1.2), and fast-and-frugal heuristics (section 1.3). From these perspectives and positions, I derived antithetical propositions, grouped them into two judgment styles (section 1.4), and tested the two styles empirically using data from a career funding scheme (sections 2 and 3).

## 1.1 The three-phase model of editorial judgment

Hirschauer (2005, 2010) introduced the three-phase model of editorial judgment based on an empirical analysis of the process used by individual editors who read and assess a manuscript before the editorial meeting. The three phases are characterized by different social frames of reference: The first phase refers to the editor's research community that shares expectations about specific types of texts or genres. The second phase refers to the editor's spontaneous impression, or judgment, that she/he has developed over the course of reading the manuscript. And the third phase represents the editor's written, rationalized judgment on the manuscript by which she/he aims to persuade the other editors. Although the three-phase model focuses on editors, I suggest that it can be generalized to the evaluation process of reviewers because the three phases are similar to the typical process of external refereeing in which peers read and assess, for example, a grant proposal or a manuscript and then provide written reviews to the editor or funding committee.

Hirschauer made several claims related to the second phase of the model that are relevant to the two research questions of this study. Specifically, Hirschauer (2004, 2015) claimed that there is a consensus on evaluation criteria among peers (i.e., in the first phase) but that this consensus quickly dissolves when criteria are practically applied (i.e., in the second phase). As his other claims related to the second phase (Hirschauer, 2004, 2015, 2019) echo Scarr (1982), I present her original position here. According to Scarr, reviewers' and editors' judgments are similar to other complex judgments, for example, those about wine, floral scents, or perfumes. Scarr argues that these judgments are based on complex weightings of many criteria and that the weights are adjusted for the criteria that apply especially to an individual case and the priorities of the referee. Accordingly, she notes that the weightings of criteria can be only partially specified and that judgments are not the result of a simple sum of component parts nor of a simple list of criteria that were combined mechanically. In line with this, Scarr (1982) refers to a referee's evaluation as a "complex human judgment" and Hirschauer (2010) as a "spontaneous impression" and a "spontaneous expression of taste". Based on Hirschauer and Scarr, the following propositions can be derived with respect to the research questions of this study: referees use different evaluation criteria; they use many criteria in their assessment; and they weight, or integrate, the criteria in a complex, non-mechanical way and on a case-by-case basis. The next two sections present theoretical perspectives and normative positions that stand in opposition to Scarr and Hirschauer.

**1.2 The ideal of impartiality and commensuration bias**

According to Lee et al. (2013), the ideal of impartiality implicitly underlies quantitative research on bias in peer review and requires that, among other things, referees "interpret and apply evaluative criteria in the same way in the assessment of a submission" so that they arrive at identical evaluations (pp. 4–5). In line with the ideal of impartiality, Forscher et al. (2019) and Arvan et al. (2022) presuppose that referees apply the same criteria when assessing the same manuscript or grant proposal. In particular, Forscher et al. (2019) argue that if referees do not agree on criteria and use different criteria, this will result in arbitrary and unreliable judgments. And Arvan et al. (2022) contend that only if referees and potential readers of a paper agree on what constitutes quality, the judgments of referees are useful to readers when deciding how to allocate their reading time among papers. In contrast to Scarr and Hirschauer, these authors state that referees are supposed to use the same evaluation criteria.

Building on the ideal of impartiality, Lee (2015) proposed commensuration bias as a new type of bias. Lee conceptualizes *commensuration* as the process by which referees convert "a submission's strengths and weaknesses for heterogeneous peer review criteria into a single metric of quality or merit" (p. 1272) and defines *commensuration bias* as referees' "deviation from the impartial weighting of peer review criteria in determinations of a submission's final value" (p. 1273). Lee argues that the deviation from impartial weighting can have three sources, one of which, referee idiosyncrasy, is relevant to the present study.[4] Referee idiosyncrasy means that referees use their own, idiosyncratic weightings and vary weightings across proposals or contexts. Idiosyncrasy is consistent with the way Scarr and Hirschauer characterize the evaluation process employed by referees (see section 1.1). According to Lee, however, referees are supposed to integrate evaluation criteria into their overall judgment with the same weightings (i.e., the same weightings as the other referees and the same weightings across proposals) to avoid commensuration bias.

**1.3 Fast-and-frugal heuristics**

Heuristics are generally defined as strategies that enable decision makers to process information in a less effortful manner than one would expect from an optimal decision rule (Shah & Oppenheimer, 2008). More specifically, in the research program on fast-and-frugal heuristics, a heuristic is defined as a procedure for making decisions under uncertainty that ignores a part of the available information in order to make decisions more quickly, frugally, and/or accurately than complex procedures (Gigerenzer et al., 2022).[5] Four concepts in this definition need further clarification: uncertainty, frugality, speed, and complex procedures. *Uncertainty* refers to "situations in which perfect foresight of all future events, their consequences, and probabilities is impossible" and in which "the optimal decision cannot be determined" (Gigerenzer et al., 2022, p. 172 and p. 174). The focus on uncertainty makes fast-and-frugal heuristics particularly useful for studying decision-making in grant funding and peer review because the outcomes of proposed research are unpredictable and judgments on research quality are associated with a high degree of uncertainty (Bornmann, 2015). *Frugality* refers to the number of cues a heuristic uses. The less cues a heuristic uses, the more frugal it is (Gigerenzer et al., 1999). In the present study, evaluation criteria represent cues. *Speed* refers to the number of operations that need to be performed to make a decision. The less computation a heuristic needs, the faster it is (Gigerenzer et al., 1999), and more frugal heuristics enable faster decisions (Wegwarth et al., 2009). *Complex procedures* assess and use all available information (i.e., they are lavish and not frugal) and integrate all information in an optimal and computationally expensive way (i.e., they are slow and not fast) (Gigerenzer & Goldstein, 1996). The literature on fast-and-frugal heuristics often mentions regression models, Bayesian models, and artificial neural networks as examples of complex procedures.

To illustrate a key finding of the fast-and-frugal literature and to show how fast-and-frugal heuristics work, I briefly describe one heuristic. The take-the-best heuristic (Gigerenzer & Goldstein, 1996) decides between two alternatives by using the cue with the highest cue validity. If no discrimination can be made, the next best cue is used, and so on. Cue validity is the accuracy with which a cue predicts a judgment (Martignon & Hoffrage, 1999). In numerous studies, the performance of this and other fast-and-frugal heuristics has been compared to the performance of complex procedures using empirical data or simulations. For example, Czerlinski et al. (1999) compared the performance of the take-the-best heuristic to the performance of multiple linear regression across 20 real-world environments using cross-validation (training, testing). They found that the heuristic was slightly more accurate in the testing set than the regression (71% versus 68% correct results), while using considerably fewer cues (2.4 versus 7.7 cues on average).[6] One of the key findings of this research is the less-is-more effect: less information and computation can lead to more accurate judgments than more information and computation (Gigerenzer et al., 2022; Gigerenzer & Goldstein, 1996). The findings from the fast-and-frugal program have the following implications for the research questions of this study: referees use a small number of the available criteria in their assessment, and they integrate the criteria into their overall judgment based on simple rules that are computationally inexpensive.

**1.4 The current study**

The antithetical propositions presented in the previous sections can be grouped into a *case-by-case* and a *uniform* style of assessment/judgment. According to the case-by-case style, referees use many and different evaluation criteria, weight criteria on a case-by-case basis, and integrate criteria in a complex, non-mechanical way into their overall judgment. "Case-by-case basis" means that referees use a distinct weighting of criteria for every proposal. According to the uniform style, referees use a small fraction of the available criteria, apply the same criteria, weight the criteria in the same way (i.e., same weightings as the other referees and the same weightings across proposals), and integrate the criteria based on simple rules (i.e., fast-and-frugal heuristics).[7] These two styles have not been examined in research on peer review, but some studies on grant funding contain evidence for the uniform style. Specifically, studies that regressed the referees' overall assessment scores on the criteria scores found that most or all criteria are positively related to the overall scores (Eblen et al., 2016; Erosheva et al., 2020; Lindner et al., 2016; Rockey, 2011; Würth et al., 2017). Moreover, a subset of these studies reported that the criteria scores explain the variability of the overall scores to a large extent (Eblen et al., 2016; Erosheva et al., 2020; Lindner et al., 2016). The results of these studies suggest that referees use the same criteria, weight them in the same way, and integrate them mechanically using a complex procedure (i.e., linear regression). However, these studies have not investigated whether referees make fast and frugal judgments (i.e., whether they use a small fraction of the available criteria and integrate the criteria based on simple rules). And the studies analyzed data from funding schemes that employed a low number of evaluation criteria (i.e., three to five), which might have concealed that referees use many and different evaluation criteria, as suggested by the case-by-case judgment style.

The current study therefore examined the case-by-case and uniform style of judgment using a unique dataset from a career funding scheme that contained a comparatively large number of evaluation criteria. A heuristic (fast-and-frugal trees) and a complex procedure (logistic regression) were employed to describe how referees integrate the criteria into their overall judgment.[8] Based on the case-by-case judgment style, I expected that a fast-and-frugal tree or regression equation could not be identified – or if they could be identified, I expected that the criteria scores would predict the overall judgment with very low accuracy – because referees use many different criteria and integrate them in a non-mechanical way and on a case-by-case basis. Based on the uniform style, I expected that a fast-and-frugal tree and regression equation could be specified and that the criteria would predict the overall judgment with high accuracy because referees use the same criteria and integrate them using the same mechanical rules and the same weightings. If a tree and regression could be specified, I expected, based on the fast-and-frugal literature, that the tree would use a small number of the evaluation criteria and deliver more accurate predictions of the referees' overall judgment than the regression that includes all evaluation criteria. In line with the fast-and-frugal paradigm, I assessed the performance of the tree and regression in a cross-validation design in which the data was split into a training set and testing set.

## 2. Methods

### 2.1 Data

The data consisted of 474 rating forms on 237 proposals that were submitted to a funding scheme for doctoral students and postdocs in two consecutive years. The purpose of the scheme is to support outstanding early career researchers from all disciplines who need funding to start or complete their research project. The unnamed funding organization relies on a small and stable pool of referees representing all disciplines to assess the applications; the referees are thus well acquainted with the evaluation procedure and the use of the rating form. From this pool, 31 referees rated the 237 proposals. The review load was unevenly distributed. Six referees each contributed more than 5% of the total reviews and accounted for 45% of all reviews; thirteen referees each contributed between 2% and 5% of all reviews, accounting for 41% of all reviews; twelve referees each contributed less than 2% of all reviews, accounting for 14% of all reviews. Every application was assessed by two referees (a first referee and a second referee). The referees rated an application on 13 criteria and provided an overall assessment (see Table 1). The majority of the evaluation criteria were scored on a five-point scale (1 poor, 2 average, 3 good, 4 very good, 5 excellent). However, the criteria "career potential" (i.e., a combination of academic potential, resilience, and long-term academic interest) and "overall assessment" included an additional scoring option (6 outstanding), while the criteria "applicant belongs to the top-5% in the field" and "proposal is highly innovative" had binary response options (yes/no). A unique feature of the present dataset is the high number of evaluation criteria, which makes it particularly suited to test the research questions. In comparison, NIH referees score proposals on five criteria (Erosheva et al., 2020), while proposals in the EU's Marie Skłodowska-Curie Actions are scored on three criteria (Pina et al., 2021), and project applications at the Swiss National Science Foundation are rated on three criteria (Würth et al., 2017).

Table 1. Descriptive statistics of the proposal ratings (training set, $n = 237$; testing set, $n = 237$).

| Evaluation criteria (cues) | Training set | | | | Testing set | | | |
|---|---|---|---|---|---|---|---|---|
| | M | SD | Mdn | min, max | M | SD | Mdn | min, max |
| Applicant | | | | | | | | |
|   Education | 4.4 | 0.7 | 4 | 2, 5 | 4.3 | 0.8 | 4 | 1, 5 |
|   Track record | 4.1 | 0.9 | 4 | 1, 5 | 4.1 | 0.9 | 4 | 1, 5 |
|   Career plan | 4.1 | 0.9 | 4 | 2, 5 | 3.9 | 1.0 | 4 | 1, 5 |
|   Career potential | 4.5 | 1.1 | 5 | 1, 6 | 4.3 | 1.0 | 4 | 2, 6 |
|   Top 5% in the field (% yes) | 14% | | | | 13% | | | |
| Project | | | | | | | | |
|   Clarity of project goal | 4.5 | 0.7 | 5 | 2, 5 | 4.5 | 0.7 | 5 | 2, 5 |
|   Clarity of research plan | 4.4 | 0.8 | 5 | 1, 5 | 4.3 | 0.9 | 5 | 2, 5 |
|   Own research question | 4.4 | 0.8 | 5 | 1, 5 | 4.2 | 1.0 | 5 | 1, 5 |
|   Methodological approach | 4.4 | 0.8 | 5 | 1, 5 | 4.3 | 1.0 | 5 | 1, 5 |
|   Feasibility | 4.2 | 0.9 | 4 | 1, 5 | 4.1 | 1.0 | 4 | 1, 5 |
|   Highly innovative proposal (% yes) | 19% | | | | 11% | | | |
| Environment | | | | | | | | |
|   Cooperation, network | 4.4 | 0.8 | 5 | 2, 5 | 4.5 | 0.7 | 5 | 2, 5 |
|   Recommendation letter | 4.4 | 0.7 | 5 | 1, 5 | 4.5 | 0.7 | 5 | 2, 5 |
| Overall assessment | 4.4 | 1.0 | 5 | 1, 6 | 4.2 | 1.2 | 4 | 1, 6 |

In the fast-and-frugal paradigm, decision rules are assessed using cross-validation to reduce overfitting and to increase generalizability (Czerlinski et al., 1999; Gigerenzer et al., 1999; Gigerenzer & Gaissmaier, 2011; Wang et al., 2022). In line with this approach, the data was split into a training set consisting of the ratings from the first referees ($n = 237$) and a testing set consisting of the ratings from the second referees ($n = 237$). The descriptive statistics of the training and testing set are shown in Table 1. As fast-and-frugal trees are designed for binary classification and decision tasks (Martignon et al., 2003; Phillips et al., 2017) and as no suitable binary dependent variable was included in the dataset, the "overall assessment" was binarized (not outstanding = scores 1 to 4; outstanding = scores 5 or 6). This binarization can be supported by two arguments. First, the scores 5 and 6 are referred to as "excellent" and "outstanding", respectively, when used as separate scores in the rating forms. However, they also have a common, superordinate meaning in the evaluation procedure and signify "absolutely necessary to be funded". Second, the funding rate of the funding scheme was 51.8% in the analyzed two years, which is almost identical to the binarized "overall assessment" of the those acting as first referee (50.6% outstanding) and somewhat higher than the binarized "overall assessment" of the those acting as second referee (45.6% outstanding). The difference of 5% between the first and second referees might be a coincidence, or it might be due to the selection procedure of the funding organization, which appoints the referee who has the most subject-matter expertise on a proposal as first referee. However, the latter explanation is unlikely as the few empirical studies on this topic suggest the opposite: the closer a referee is to a proposal, the harsher the evaluations (for a discussion, see Gallo et al., 2016).

**2.2 Statistical analysis**

Many fast-and-frugal heuristics are available, and this study employed one particular heuristic, fast-and-frugal-trees, for the following reasons. Unlike the class of one-clever-cue heuristics (Gigerenzer et al., 2022), fast-and-frugal trees are flexible in terms of the number of cues included because they can be created with just one cue or expanded to include more cues (Martignon et al., 2003). They are therefore suited to examine the two judgment styles proposed in this study. Moreover, fast-and-frugal trees are lexicographic and thus non-compensatory, that is, the cues are used in order of their cue validity and later cues thus cannot reverse the decision made by an earlier cue (Gigerenzer et al., 1999). For example, if a grant proposal is considered unoriginal (first cue) and therefore rejected, its methodological strength (second cue) or feasibility (third cue) cannot compensate for this because later cues remain unconsidered in a fast-and-frugal tree. In contrast, complex procedures such as regression models can compensate unoriginality by methodological rigor or feasibility because they integrate all cues. Fast-and-frugal trees thus use a markedly different type of integration logic than complex procedures, which might be useful for understanding peer review judgments in a novel way. Lastly, fast-and-frugal trees have been studied in a variety of contexts, including medical, legal, financial, and managerial decision-making (Phillips et al., 2017), but not in the domain of peer review.

Fast-and-frugal trees consist of cues sequentially ordered by cue validity (i.e., from high to low) and binary decisions based on these cues (Martignon et al., 2003, 2008). A tree can be represented graphically as a decision tree or described verbally by a series of if-then-else statements. For example, if a grant proposal is considered unoriginal then reject it; otherwise, examine whether it is methodologically sound. If it is not sound, reject it; otherwise, examine whether it is feasible. If it is not feasible, reject it; otherwise, fund it. In the current study, fast-and-frugal trees were constructed and assessed in R 4.1.2 (R Core Team, 2021) using the FFTrees package 1.6.6 (Phillips et al., 2017). The training set was used to fit fast-and-frugal trees with one to six cues (six is the maximum number of cues the FFTrees package can fit). The 13 evaluation criteria served as cues and the binarized "overall assessment" as criterion variable. In line with the requirement to ignore conditional dependencies between cues when building fast-and-frugal trees (Martignon et al., 2003), the ifan algorithm was used to create the trees. From the trees created by the ifan algorithm, those with the highest accuracy were selected. For cross-validation, the resulting six fast-and-frugal trees were then applied to the testing set and their performance was computed (see performance metrics below).

This study employed logistic regression as a complex procedure because regression models are a popular method and benchmark in studying human judgment, including peer review. For example, regression models have been used for decades to directly study and describe human judgment and behavior in a variety of contexts (Beckstead, 2007; Gigerenzer et al., 2011), including peer review (e.g., Lindner et al., 2016; Porter & Rossini, 1985; Reinhart, 2009). Moreover, regression models are often used as benchmark in fast-and-frugal studies (Gigerenzer et al., 2011) and sometimes serve as benchmark in peer review studies (e.g., Devyatkin et al., 2018; Kang et al., 2018).

The training set was used to fit a logistic regression. Specifically, the binarized "overall assessment" was regressed on the 13 evaluation criteria using the glm function and the binomial family implemented in R 4.1.2 (R Core Team, 2021). The cutoff probability to classify the overall judgment as "outstanding" and "not outstanding" was determined using the argument "misclasserror" in the optimalCutoff function from the InformationValue package 1.2.3 (Prabhakaran, 2016), which minimizes the misclassification rate. Collinearity was assessed exploiting the vif function from the car package 3.1.0 (Fox & Weisberg, 2019). For cross-validation, the resulting regression model was then applied to the testing set and its performance was computed (see performance metrics below). In addition to regressing all 13 criteria on the overall judgment, two regressions with the most important criteria only were computed to assess how "frugal" regressions perform. The most important criteria were selected in two ways. On the one hand, the same criteria were used in the regression as in the best performing fast-and-frugal tree (i.e., the criteria with the highest cue validities). On the other hand, the "significant" criteria from the full regression were included using a traditional frequentist statistics cutoff of $p = 0.05$ (i.e., the criteria with the highest log-odds). The two regressions were computed in the training set, and the resulting models were then applied to the testing set.

The performance of the trees and regression models was evaluated based on five metrics. *Absolute frugality* (#Frug) is the average number of cues a decision rule integrates to reach a decision. For example, a heuristic that consists of two cues and uses one cue to reach a decision in 40% of all decisions and two cues in 60% of all decisions has an absolute frugality of 1.6. *Relative frugality* (%Frug) represents the share of cues that are ignored and equals 1 minus the absolute frugality divided by all available cues. For example, a heuristic ignores 88% of the cues if 13 cues are available and the heuristic uses 1.6 cues on average.[9] In addition to the frugality metrics, three metrics based on the confusion matrix were used. *Accuracy* (Acc) is the sum of true positives and true negatives divided by the number of all cases or decisions. It is the most widely used metric in research on fast-and-frugal heuristics. *Sensitivity* (Sens) is the number of true positives divided by the sum of true positives and false negatives. *Specificity* (Spec) is the number of true negatives divided by the sum of true negatives and false positives.

## 3. Results

The cue validities of all thirteen evaluation criteria, which were calculated from the ratings in the training set ($n = 237$), are shown in Figure 1. Cue validity is defined as the accuracy with which a cue predicts a judgment; a cue is informative and valid for making a judgment if its accuracy is > 0.5 (Martignon & Hoffrage, 1999). According to Figure 1, the accuracy of all cues is > 0.5, which indicates that all cues are valid and, more broadly, that all evaluation criteria are related to the overall judgment. The "career potential" of the applicant has the highest cue validity (0.87), while the criterion "applicant belongs to the top-5% in the field" has the lowest cue validity (0.63). Figure 2 shows fast-and-frugal-trees describing the referees' overall judgments in the training set ($n = 237$) with one, two, and three cues (i.e., evaluation criteria). According to these trees, the "career potential" of the applicant is the most important evaluation criterion for referees, which is consistent with the goal of the funding scheme to promote outstanding early career researchers. The two other evaluation criteria, "methodological approach" and "letter of recommendation", are placed at lower levels of the trees due to their lower cue validity.

Figure 1. Cue validities of the thirteen evaluation criteria in the ROC space (training set, *n* = 237).

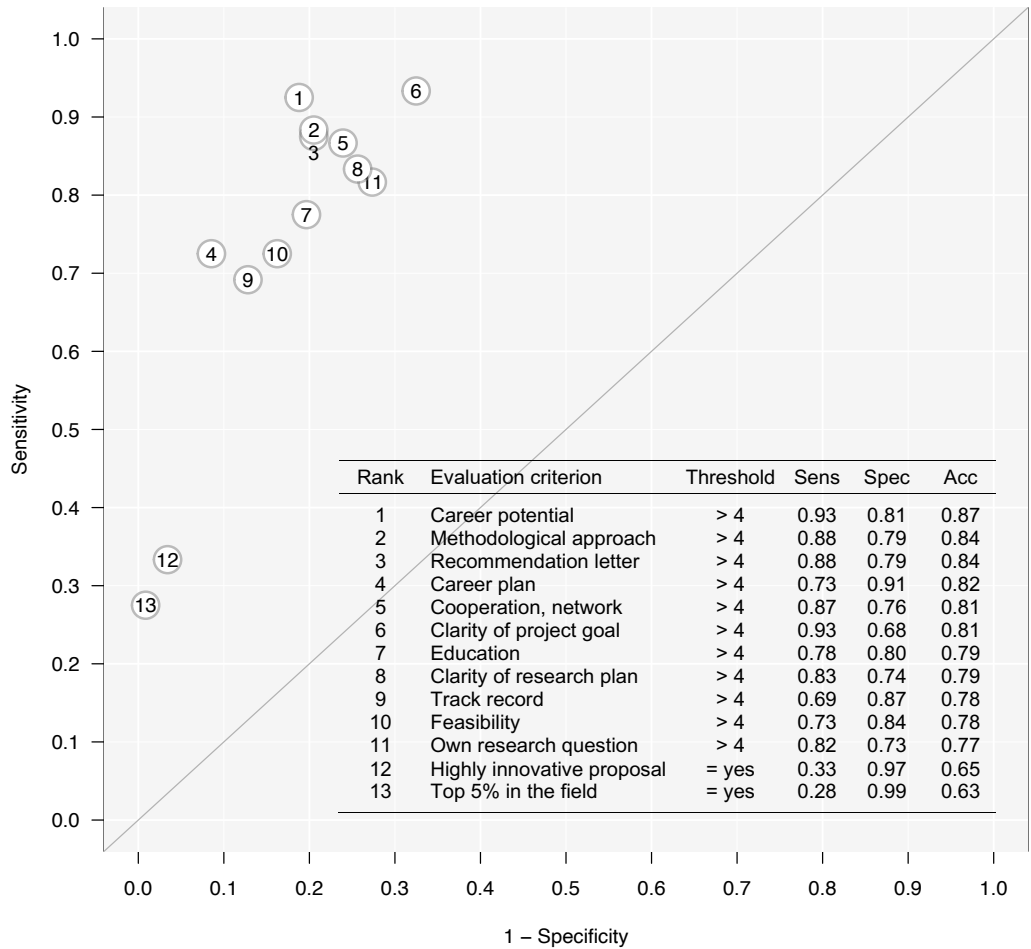

| Rank | Evaluation criterion | Threshold | Sens | Spec | Acc |
|---|---|---|---|---|---|
| 1 | Career potential | > 4 | 0.93 | 0.81 | 0.87 |
| 2 | Methodological approach | > 4 | 0.88 | 0.79 | 0.84 |
| 3 | Recommendation letter | > 4 | 0.88 | 0.79 | 0.84 |
| 4 | Career plan | > 4 | 0.73 | 0.91 | 0.82 |
| 5 | Cooperation, network | > 4 | 0.87 | 0.76 | 0.81 |
| 6 | Clarity of project goal | > 4 | 0.93 | 0.68 | 0.81 |
| 7 | Education | > 4 | 0.78 | 0.80 | 0.79 |
| 8 | Clarity of research plan | > 4 | 0.83 | 0.74 | 0.79 |
| 9 | Track record | > 4 | 0.69 | 0.87 | 0.78 |
| 10 | Feasibility | > 4 | 0.73 | 0.84 | 0.78 |
| 11 | Own research question | > 4 | 0.82 | 0.73 | 0.77 |
| 12 | Highly innovative proposal | = yes | 0.33 | 0.97 | 0.65 |
| 13 | Top 5% in the field | = yes | 0.28 | 0.99 | 0.63 |

Note: Acc = accuracy; Sens = sensitivity; Spec = specificity.

The tree with two cues (Figure 2, b) indicates that a referee assesses a proposal as outstanding only if the applicant's career potential is excellent or outstanding (scores 5 and 6, respectively) and the methodological approach is excellent (score 5). The tree thus represents a conjunctive decision rule and should minimize false positives (Einhorn, 1970; Gigerenzer et al., 2022). In fact, the tree demonstrates high specificity and lower sensitivity (Table 3) and generated a total of 14 false positives and 49 false negatives (*N* = 474). In contrast, the tree with three cues (Figure 2, c) represents a zigzag pattern and should thus minimize both false positives and false negatives (Gigerenzer et al., 2022; Martignon et al., 2003). As Table 3 demonstrates, sensitivity and specificity are more balanced in this tree than in the two-cue tree. More specifically, the tree with three cues produced a total of 29 false positives and 37 false negatives (*N* = 474). The tree indicates that a referee does not evaluate an application unfavorably if the methodological approach is not sufficient (scores 1 to 4), but she/he considers further information, the letter of recommendation, to make the final judgment (Figure 2, c). As research on fast-and-frugal heuristics prefers results from the testing set to results from the training set due to greater generalizability (Gigerenzer et al., 1999; Wang et al., 2022), the tree with two cues in the testing set and an accuracy of 0.85 can be considered the one that best describes the judgment process of referees although it is only marginally more accurate than the other trees in the testing set (Table 3). This tree uses on average 1.4 cues in the testing set to predict the referees' overall judgment and thus ignores 89% of the available cues (training set: #Frug = 1.6, %Frug = 0.88). A verbal description of the trees consisting of four, five, and six cues is provided in the Appendix (Table A1).

Figure 2. Fast-and-frugal-trees describing the referees' overall judgment (outstanding, not outstanding) with one, two, and three cues (training set, $n = 237$).

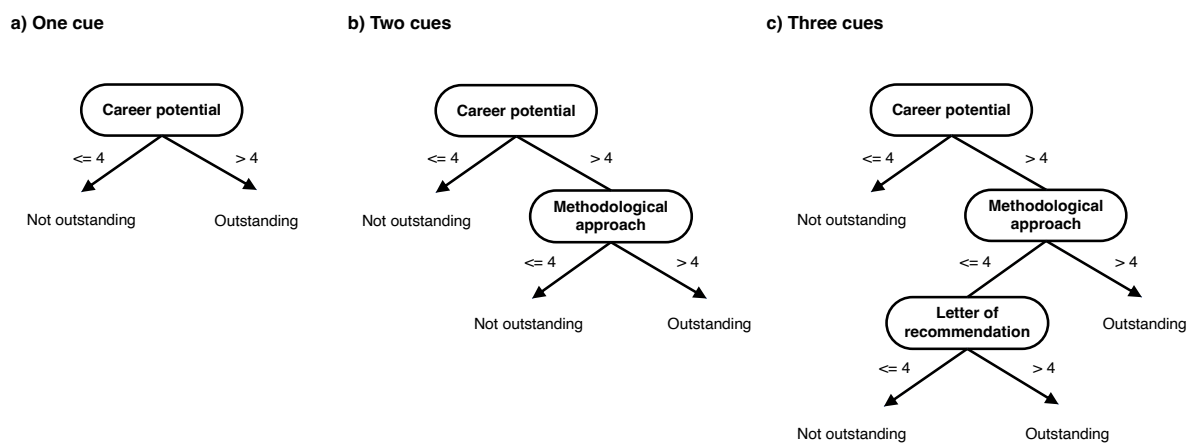

Table 2 shows the results of the logistic regression, in which the referees' overall judgments in the training set ($n = 237$) were regressed on all thirteen evaluation criteria (AIC = 123.95; $X^2$ (13, $N = 237$) = 232.56, $p < 0.001$; pseudo-$R^2$: McKelvey-Zavoina = 0.90, Veall-Zimmermann = 0.85, McFadden = 0.71). According to the estimates in Table 2, all evaluation criteria are positively related to the overall judgment, suggesting that referees use a broad range of criteria to evaluate an application and integrate the criteria in a linear, additive way. However, the confidence intervals of the estimates are wide, likely due to collinearity. Rules of thumb for identifying collinearity suggest that a VIF < 4, < 5, or even < 10 indicates low collinearity. The VIFs reported in Table 2 could thus be considered unproblematic. Johnston et al. (2018), however, argued that even low levels of collinearity (VIF < 2.5) can lead to unreliable regression coefficients. Moreover, Lindner et al. (2016) regressed the overall assessment of NIH referees on five evaluation criteria and found that the correlated criteria render regression coefficients unreliable. Caution should therefore be exercised when determining the importance of individual evaluation criteria based on the log-odds and p-values reported in Table 2. Although Table 2 does not provide conclusive evidence that referees use all criteria, the estimates, the superior accuracy (Table 3), and the better goodness-of-fit statistics of the regression model including all cues (see statistics of the "frugal" regressions below) suggest that referees do so.

Table 2. Logistic regression analysis of the referees' overall judgments (training set, $n = 237$).

| Variable | Estimate | SE | 95% CI LL | 95% CI UL | p | VIF |
|---|---|---|---|---|---|---|
| Applicant | | | | | | |
|     Education | 0.26 | 0.57 | -0.87 | 1.40 | .66 | 2.2 |
|     Track record | 1.42 | 0.47 | 0.55 | 2.40 | <.01 | 2.6 |
|     Career plan | 0.53 | 0.43 | -0.31 | 1.38 | .21 | 2.7 |
|     Career potential | 0.87 | 0.49 | -0.07 | 1.86 | .07 | 3.8 |
|     Top 5% in the field | 0.73 | 1.38 | -1.63 | 4.13 | .60 | 1.4 |
| Project | | | | | | |
|     Clarity of project goal | 1.51 | 0.68 | 0.24 | 2.93 | .03 | 2.4 |
|     Clarity of research plan | 0.69 | 0.54 | -0.36 | 1.79 | .21 | 2.6 |
|     Own research question | 0.49 | 0.52 | -0.55 | 1.52 | .35 | 2.1 |
|     Methodological approach | 1.01 | 0.54 | -0.04 | 2.09 | .06 | 2.9 |
|     Feasibility | 0.09 | 0.41 | -0.74 | 0.89 | .83 | 2.1 |
|     Highly innovative proposal | 0.50 | 0.74 | -0.90 | 2.06 | .50 | 1.3 |
| Environment | | | | | | |
|     Cooperation, network | 0.34 | 0.53 | -0.69 | 1.40 | .52 | 2.3 |
|     Recommendation letter | 0.82 | 0.46 | -0.08 | 1.74 | .07 | 2.2 |
| Constant | -36.15 | 5.99 | -49.49 | -25.81 | <.01 | |

Note: CI = confidence interval; Estimate = log-odds; SE = standard error; LL = lower limit; UL = upper limit; VIF = variance-inflation factor.

Table 3 indicates that the regression that includes all thirteen cues is slightly more accurate than the fast-and-frugal tree that best describes the referees' judgment process (i.e., the two-cue tree). More specifically, the regression outperforms the tree in terms of accuracy in both training (0.92 vs 0.88) and testing (0.89 vs 0.85). However, in contrast to the tree, the regression is lavish because it uses all available cues (#Frug = 13) and ignores none (%Frug = 0). Hence, much more information and computation produced slightly more accurate judgments than considerably less information and computation.

To assess how "frugal" regressions would perform, two regressions with statistically significant criteria only (referred to as log-odds model) and the criteria from the best performing fast-and-frugal tree (referred to as cue-validity model) were computed. Although it was pointed out above that interpreting the log-odds and p-values is probably not very meaningful, a frugal regression model was defined based on statistical significance because blindly using statistical significance is a typical strategy employed in judgment analysis studies to identify the cues that judges use (Beckstead, 2007). When applying this strategy and using the traditional frequentist statistics cutoff of $p = 0.05$, the two criteria with the highest log-odds in Table 2, "clarity of the project goal" ($p = 0.002$) and "track record" ($p = 0.03$), would be considered the cues that the referees have used. This model, however, is less accurate than the cue-validity model, which included the criteria from the best performing fast-and-frugal tree (i.e., "career potential" and "methodological approach", the cues with the highest cue validities). More specifically, the log-odds model is less accurate than the cue-validity model in both training (0.85 vs 0.89) and testing (0.82 vs 0.85; Table 3). This is also reflected in the goodness-of-fit statistics of the log-odds model (AIC = 161.94; $X^2$ (2, $N$ = 237) = 172.57, $p < 0.001$; pseudo-$R^2$: McKelvey-Zavoina = 0.78, Veall-Zimmermann = 0.73, McFadden = 0.53), which are less favorable than those of the cue-validity model (AIC = 142.09; $X^2$ (2, $N$ = 237) = 192.42, $p < 0.001$; pseudo-$R^2$: McKelvey-Zavoina = 0.81, Veall-Zimmermann = 0.77, McFadden = 0.59).

The two frugal regression models provide a baseline against which the performance of the regression including all cues can be assessed. While the frugal models both ignore 85% of the available cues, their accuracy is not considerably lower than that of the model that includes all cues and ignores none (Table 3). In particular, the log-odds and cue-validity model are only slightly less accurate than the model including all cues in both training (0.85 and 0.89 vs 0.92) and testing (0.82 and 0.85 vs 0.89). Using significantly more criteria in a complex procedure thus increased accuracy only slightly.

Since the cue-validity model and the two-cue tree consist of the same criteria (i.e., career potential, methodological approach), the performance of the two decision rules, logistic regression and fast-and-frugal trees, can be directly compared. According to Table 3, the two-cue tree is slightly more frugal than the cue-validity model because it ignores 88% (training) and 89% (testing) of the available cues, whereas the regression ignores 85% of the cues. The tree, however, is much faster (i.e., computationally less demanding) because it performs a few simple comparisons (i.e., "is less than", "is greater than") to make a decision, while the logistic regression needs to execute multiplications and additions plus a comparison with a cutoff. In terms of accuracy, the tree and the regression are virtually identical in training (0.882 vs 0.890) and testing (0.852 vs 0.848; the respective values are rounded to two decimals in Table 3). Hence, when the same few evaluation criteria were used, fast-and-frugal trees offered a more parsimonious explanation of the referees' overall judgment than logistic regression.

Table 3. Performance of fast-and-frugal trees and logistic regressions in the training set (237 judgments) and testing set (237 judgments).

| Decision rule | Training set | | | | | Testing set | | | | |
|---|---|---|---|---|---|---|---|---|---|---|
| | #Frug | %Frug | Acc | Sens | Spec | #Frug | %Frug | Acc | Sens | Spec |
| Fast-and-frugal trees | | | | | | | | | | |
| 1 cue | 1 | 92% | .87 | .93 | .81 | 1 | 92% | .82 | .80 | .84 |
| 2 cues | 1.6 | 88% | .88 | .83 | .94 | 1.4 | 89% | .85 | .74 | .95 |
| 3 cues | 1.7 | 87% | .90 | .90 | .91 | 1.5 | 88% | .82 | .77 | .86 |
| 4 cues | 1.7 | 87% | .90 | .88 | .92 | 1.6 | 88% | .84 | .77 | .91 |
| 5 cues | 1.7 | 87% | .91 | .89 | .92 | 1.6 | 88% | .84 | .77 | .90 |
| 6 cues | 1.8 | 86% | .91 | .89 | .92 | 1.6 | 88% | .84 | .77 | .90 |
| Logistic regression | | | | | | | | | | |
| 2 cues (log-odds) | 2 | 85% | .85 | .91 | .79 | 2 | 85% | .82 | .91 | .75 |
| 2 cues (cue validity) | 2 | 85% | .89 | .86 | .92 | 2 | 85% | .85 | .74 | .94 |
| All cues | 13 | 0% | .92 | .95 | .89 | 13 | 0% | .89 | .88 | .90 |

Note: #Frug = absolute frugality; %Frug = relative frugality; Acc = accuracy; Sens = sensitivity; Spec = specificity.

To assess the robustness of the performance of the fast-and-frugal trees and logistic regressions reported in Table 3, a second training and testing set was created using stratified random sampling. Specifically, the training set contained 50% of the judgments from the first referees and 50% of the judgments from the second referees (total 238 judgments). The testing set contained the remaining 236 judgments (50% from the first referees and 50% from the second referees). The results are provided in the Appendix (Figure A1). They are consistent with the results in Table 3 and lead to the same conclusions.

To examine whether only a sub-group of the 31 referees used the same decision rule, for example those referees contributing most reviews, and thus account for the results of this study, the share of a referee's total judgments correctly described by the fast-and-frugal tree (2 cues) and the logistic regression (13 cues) were calculated. Figure 3 (training set) and Figure 4 (testing set) show that the tree and the regression describe a large proportion of each referee's judgments and that the regression's performance is superior to that of the tree. More specifically, the tree correctly describes 80% or more of the judgments of 24 referees in the training set and of 22 referees in the testing set, while the regression correctly describes 80% or more of the judgments of 26 referees in both the training and testing set. Moreover, the regression describes the judgments of eleven referees more accurately than the regression in the training set (referees 2, 3, 5, 6, 7, 8, 11, 12, 16, 19, 25) and of seven referees in the testing set (referees 2, 3, 5, 13, 23, 25, 31). In contrast, the tree describes the judgments of only four referees more accurately than the regression in the training set (referees 1, 10, 14, 26) and of three referees in the testing set (referees 1, 14, 18). In Figures 3 and 4, the variability of the correctly described judgments increases from left (referees contributing the most reviews) to right (referees contributing the least reviews), which is likely due to the decreasing sample size. Overall, Figures 3 and 4 suggest that referees mostly use the same decision rule as their fellow referees and the same rule across proposals, providing support for the uniform judgment style. However, Figures 3 and 4 also clearly demonstrate that not all referees apply the same decision rule all the time, which is logically consistent with the high but not perfect accuracy values reported in Table 3.

Figure 3. Training set: Proportion of a referee's total judgments correctly described by the two-cue fast-and-frugal tree and the logistic regression that included all thirteen cues (referees: $n = 30$, judgments: $n = 237$).

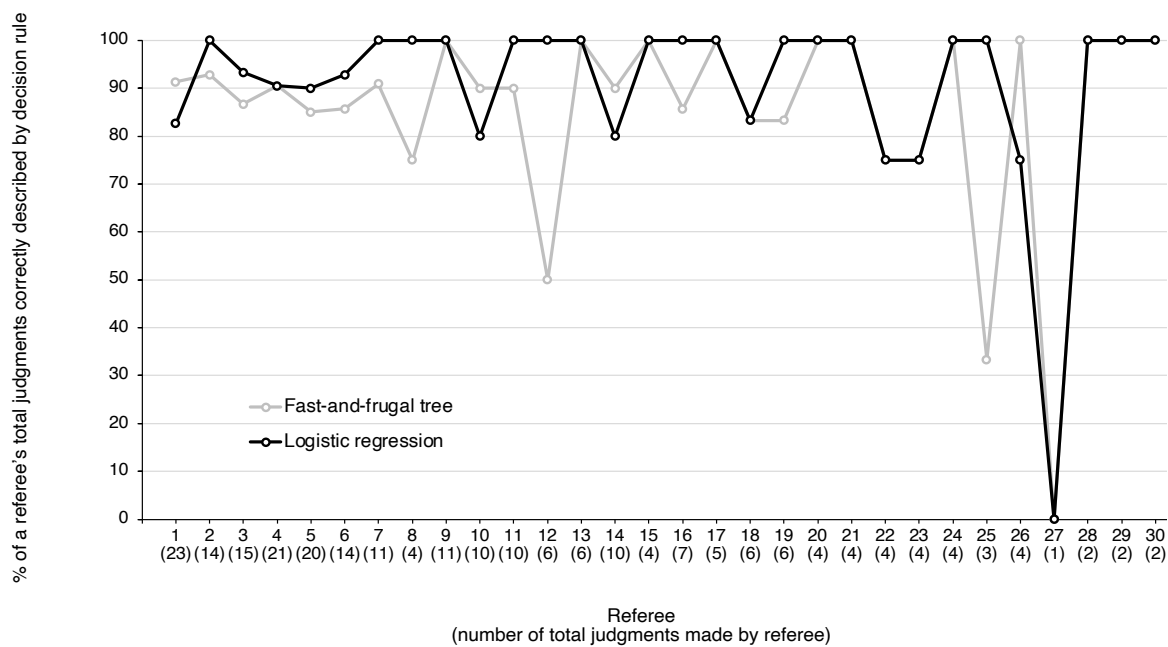

Figure 4. Testing set: Proportion of a referee's total judgments correctly described by the two-cue fast-and-frugal tree and the logistic regression that included all thirteen cues (referees: $n = 31$, judgments: $n = 237$).

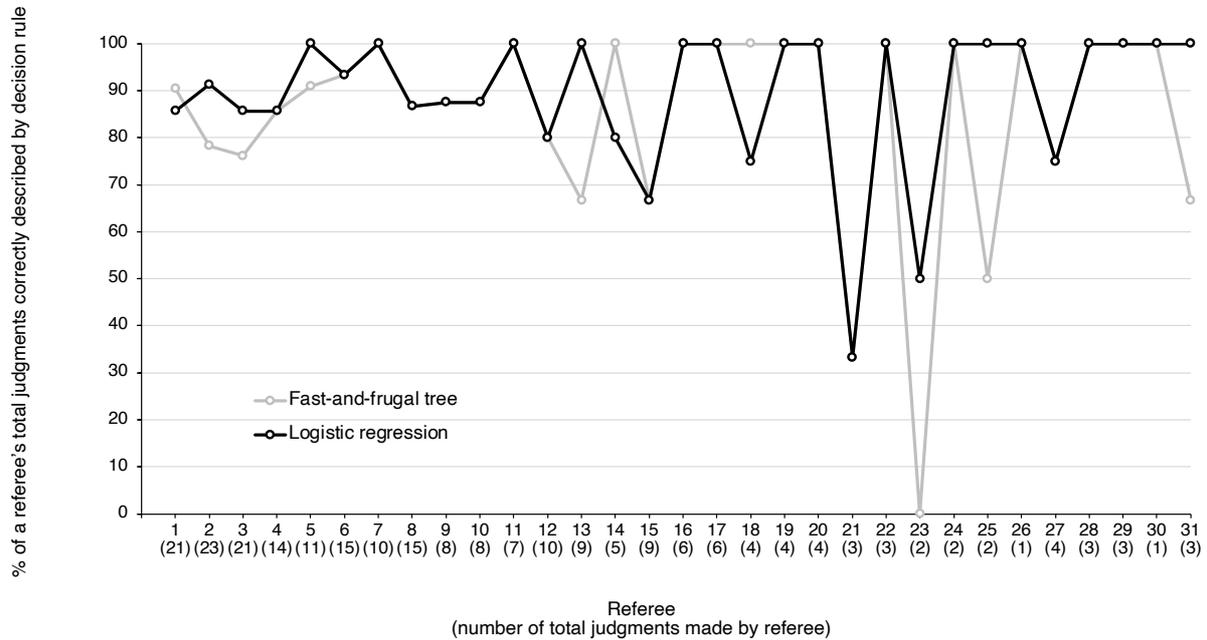

## 4. Discussion and conclusion

The evaluation process that referees employ in academic peer review is undertheorized. With respect to this process, little is known whether referees use the same evaluation criteria and how they integrate the criteria into their overall judgment. This study therefore proposed two assessment styles, the case-by-case and the uniform style, based on theoretical perspectives and normative positions. These styles were tested using data from a career funding scheme for doctoral students and postdocs from all disciplines. The results of this study suggest that referees use many evaluation criteria, apply the same criteria, weight the criteria in the same way (i.e., the same weightings as the other referees and the same weightings across proposals), and integrate the criteria mechanically using a complex rule (i.e., a linear-additive model). This is generally consistent with the uniform judgment style. But, based on the results of this study, the uniform style needs to be revised in two respects. The uniform style proposed that referees use a fraction of the available criteria, while the results suggest that referees use *many criteria*.

In addition, the *complex procedure* (logistic regression) was more accurate than the simple rule (fast-and-frugal heuristics) proposed in the uniform style. However, and most importantly, the revised style applies to most – but not all – of the referees' judgments in this study. The fact that the revised uniform style cannot describe all judgments could be due to the following reasons. First, the referees sometimes used a case-by-case judgment style, that is, they weighted the criteria on a case-by-case basis, applied different criteria, or integrated the criteria using other rules. Second, the decision rules employed in this study (logistic regression, fast-and-frugal trees) were unable to reveal the true rules referees apply. Other heuristics or other complex algorithms, such as elastic net regression or support-vector machines, may be better suited to capture how referees integrate evaluation criteria. Lastly, the referees uniformly used additional criteria not covered by the rating form. For example, referees may have systematically looked up the citation impact of the applicants in bibliometric databases or based their judgment also on bias factors, such as status or gender. Or referees applied universalistic merit criteria that were not included in the rating form.

The results from the logistic regression including all cues are consistent with studies on grant peer review that regressed the referees' overall assessment scores on the criteria scores. Similar to the present study, these studies found that most or all evaluation criteria are positively related to the referees' overall scores (Eblen et al., 2016; Erosheva et al., 2020; Lindner et al., 2016; Rockey, 2011; Würth et al., 2017) and that the criteria scores explain the variability of the overall scores to a large extent but not completely (Eblen et al., 2016; Erosheva et al., 2020; Lindner et al., 2016). These studies, however, used data from funding schemes in which referees scored the proposals on three to five criteria, while the present study analyzed data from a funding instrument that encompassed thirteen criteria. The present study thus extends the results of previous research and suggests that referees use many criteria in their assessment. This finding, which refers to the second phase in Hirschauer's (2005, 2010) three-phase model of judgment (i.e., the judgment developed over the course of reading an application), is in line with results from studies that focused on the first phase (i.e., the expectations shared in a research community) and the third phase (i.e., the written post hoc rationalization of the judgment). More specifically, a systematic review identified a broad set of criteria peers refer to in the evaluation of grant applications (Hug & Aeschbach, 2020), and empirical studies demonstrated that scholars' notions of research quality and performance are multifaceted (Andersen, 2013; Bazeley, 2010; Gulbrandsen, 2000; Hemlin, 1993; Hemlin & Montgomery, 1990; Hug et al., 2013; Margherita et al., 2022; Mårtensson et al., 2016; Ochsner et al., 2013; Prpić & Šuljok, 2009). Moreover, the finding is in line with Hren et al. (2022) who showed that almost all of the 29 themes (or criteria) identified in the evaluation summaries written by grant panels are related to the panels' decisions.

The results of this study do not support one of the key findings of the research on fast-and-frugal heuristics, the less-is-more effect, which states that less information and computation can lead to more accurate judgments than more information and computation (Gigerenzer et al., 2022; Gigerenzer & Goldstein, 1996). In accordance with the less-is-more effect, Raab and Gigerenzer (2015) as well as Phillips et al. (2017) emphasize that fast-and-frugal trees can be as accurate as or more accurate than complex procedures. Empirical studies, however, show a more nuanced picture. Fast-and-frugal trees were found to be more accurate than (e.g., Dhami & Ayton, 2001; Wegwarth et al., 2009), as accurate as (e.g., Aikman et al., 2021; Jenny et al., 2013), or less accurate than (e.g., Woike et al., 2015) models integrating all cues. This range of results was also obtained by Phillips et al. (2017) who compared fast-and-frugal trees to six complex procedures using ten real-world datasets. They found that, in the testing set, the accuracy of fast-and-frugal trees (0.83) was higher than that of naïve Bayes (0.78) and CART (0.80), comparable to logistic regression (0.82), regularized logistic regression (0.83), and random forests (0.83), but lower than that of support-vector machines (0.86). Moreover, several studies demonstrated that when the size of the dataset for training was small (i.e., 15% of the total data), fast-and-frugal trees tended to do best, producing either comparable or slightly better results in the testing set than complex procedures (Laskey & Martignon, 2014; Martignon et al., 2008, 2012; Woike et al., 2017). When 50% of the data was included in the training set, logistic regression was 2 to 3% more accurate than fast-and-frugal trees (Martignon et al., 2008, 2012), which corresponds to the design and the results of the current study. The results of the present study are thus in line with previous studies on fast-and-frugal trees but do not support the less-is-more effect.

The study has the following main limitations. First, it focused on a funding scheme for doctoral students and postdocs, and it analyzed data from a relatively small pool of referees. It thus remains to be investigated whether the findings can be generalized to a larger population of referees, other funding schemes, and other types of academic peer review. Second, the study examined the performance of only two decision rules (fast-and-frugal trees, logistic regression) and thus may not have considered the rule that can most accurately describe referees' judgments. Future studies may therefore include a broader range of heuristics and complex procedures. Third, the analysis focused on evaluation criteria provided by the funding organization and did not include characteristics of the applications, applicants, or referees (e.g., gender, age, discipline). The effect of potential bias factors has thus not been assessed. Future studies should therefore include bias factors, such as gender (Cruz-Castro & Sanz-Menendez, 2021; Sato et al., 2021; Schmaling & Gallo, 2023; Squazzoni et al., 2021), when examining the two judgment styles. Fourth, the referees' ratings were analyzed using a nomothetic, group-level approach (Beltz et al., 2016; Piccirillo & Rodebaugh, 2019), which may have favored the uniform judgment style. Future research may therefore employ an idiographic approach and examine the judgments of each referee individually. Fifth, the ratings of the proposals were treated as events independent of the referees. As grant peer review is inherently multilevel (Erosheva et al., 2020), future research may use multilevel approaches to examine the two judgment styles proposed in this study. Lastly, the study was based on the assumption that criteria scores reported on rating forms allow inferences to be made about the referees' evaluation process and the two judgments styles. However, data collected during the assessment process may be more appropriate for examining the two styles. For example, Vallée-Tourangeau et al. (2022) used the think aloud method to explore factors influencing the evaluation of grant applications.

The results of this study, if generalizable, have several implications for peer review research and practice. First, the results support the practice of many funding agencies to score applications on a small number of evaluation criteria. According to Langfeldt and Scordato (2016), funding agencies use few criteria for reasons of simplicity, clarity, flexibility, and efficiency. The present study demonstrated that as few as two evaluation criteria can explain the variability of the referees' overall judgments to a large extent, whereas including significantly more criteria improved the explained variation only slightly. This suggests that funding agencies using few criteria sacrifice little quantitative information in exchange for greater simplicity, clarity, and flexibility. Nevertheless, funders must be aware that referees use many evaluation criteria. To complement the quantitative criteria in the rating forms, it is therefore reasonable to retain the written assessments widely used by funding agencies. Moreover, funding agencies that do not ask referees to provide an overall score but calculate overall scores from a small number of criteria or agencies that base their funding decisions solely on criteria scores need to be aware that they ignore some of the evaluative information referees can provide and thus likely alter the results of the peer review process. Second, for modeling peer review judgments, fast-and-frugal trees can represent an alternative to more complex statistical methods, depending on the research context. When only few data are available or collecting data is costly, fast-and-frugal trees provide an effective alternative because trees are relatively robust against overfitting (Phillips et al., 2017) and produce comparable or slightly better results than complex procedures (Laskey & Martignon, 2014; Martignon et al., 2008, 2012; Woike et al., 2017). In addition, fast-and-frugal trees have several advantages that can outweigh the slight loss in accuracy found in this and other studies. Trees typically use very little information (i.e., cues); they are computationally inexpensive; and they enable fast decisions. They are easy to understand, communicate, and apply because they are simple and transparent (Phillips et al., 2017). And trees, like other fast-and-frugal heuristics, are designed for studying decision-making situations characterized by uncertainty (Gigerenzer et al., 2022); peer review represents such a situation. Third, the present study has shown that referees mostly use the same criteria and integrate the criteria in the same way. This suggests that the low inter-rater reliability consistently observed in peer review (Bornmann, 2011; Bornmann et al., 2010; Cicchetti, 1991; Erosheva et al., 2021; Lee et al., 2013) is less likely to be caused by referees using different criteria and integration rules.

As this study presented evidence that supports the uniform judgment style, future studies should attempt to falsify this judgment style. Future studies should also examine how referees' judgments can be characterized in those cases where the uniform style fails to provide an accurate and proper description. Lastly, and more broadly, peer review research should study the evaluation process of referees in more empirical and theoretical detail. This would advance our understanding of how peer review works and how we can develop it further.

## Acknowledgments

I am grateful to Anna Ekert-Centowska for her thoughtful comments and suggestions on earlier versions of this paper. I would like to extend my sincere thanks to Rüdiger Mutz for helpful discussions. I also wish to thank the reviewers for thoroughly engaging with the manuscript and for providing constructive feedback.– 22 –

**Funding information**

No funding was received for conducting this study.

**Competing interests**

The author has no competing interests.

**Data availability**

Data is not available due to legal restrictions.

**Notes**

1. The central role of peer review has been re-emphasized in a widely supported European initiative to reform research assessment practices (European Commission, 2021; CoARA, 2022).
2. There are many other questions on which we have made little progress, for example: How can we compare and contrast the many facets of peer review across contexts (disciplines, purposes, regions, etc.)? What are the purposes and roles of peer review, and how do peer review processes (not) achieve them? How can we define and conceptualize peer review, given that it is a highly diverse practice, is institutionalized in various ways, includes many different procedures, serves different purposes, and evolves through time? What are the most pressing challenges for peer review systems, what are the causes of these challenges, and which peer review innovations can effectively address these challenges?
3. I use the terms *reviewer* and *referee* interchangeably. A referee is a researcher who contributes an independent and external evaluation of the work of fellow researchers (e.g., a research proposal, a registered report, a manuscript) to a peer review process.
4. In addition to referee idiosyncrasy, Lee (2015) mentions two other sources for commensuration bias. First, referees weight criteria according to social characteristics of the applicants or authors. Second, referees undervalue criteria that promote truth and innovation in science (e.g., methodological soundness, novelty).
5. The research program on fast and frugal heuristics should not be confused with the *heuristics and biases approach* (Gilovich et al., 2002; Tversky & Kahneman, 1974), which, in contrast to the fast-and-frugal program, favors a skeptical attitude towards human judgment (Kahneman & Klein, 2009) and has identified heuristics as a source of biases and errors (Gilovich et al., 2002; Tversky & Kahneman, 1974).

6. The results, however, were the other way around in the training set: the regression was slightly more accurate (77% correct) than the heuristic (75% correct). This very pattern is often observed in studies on fast-and-frugal heuristics: complex procedures overfit, while heuristics avoid overfitting due to their simplicity and thus outperform complex procedures in the testing set (Artinger et al., 2022; Gigerenzer et al., 1999, 2011). The robustness and generalizability of fast-and-frugal heuristics is thus often higher than that of complex procedures.

7. The case-by-case style and the uniform style resemble Meehl's distinction between clinical and statistical judgment (Meehl, 1954). Note that the case-by-case and uniform style are both conceived as human judgments, while Meehl only sees clinical judgment as a judgment that is formed "in the head" of humans (Grove & Meehl, 1996, p. 293). From Meehl's perspective, both styles would be considered clinical because they are formed by humans (Meehl, 1954; Grove, 2005).

8. While the present study is the first that models peer review judgments using fast-and-frugal heuristics, fast-and-frugal heuristics have already been employed in the domain of research evaluation to describe and prescribe the use of bibliometric indicators (e.g., Bornmann et al., 2022; Bornmann & Hug, 2020; Bornmann & Marewski, 2019; de Abreu Batista Júnior et al., 2021).

9. I use the same two metrics as Phillips et al. (2017) (i.e., mean cues used; percent cues ignored) but name them differently (i.e., absolute frugality; relative frugality). Moreover, I interpret the metric "mean cues used" differently. While Phillips and colleagues refer to "mean cues used" as a speed measure, I interpret it as a measure of frugality (i.e., absolute frugality), which is line with Czerlinski et al. (1999) who defined frugality as the "average number of cues looked up" (p. 103).

# Appendix

Table A1. Verbal description of fast-and-frugal-trees describing referees' overall judgment with four, five, and six cues.

| No. of cues | Description |
|---|---|
| Four cues | Like the three-cue tree (Figure 2, c) but replace its lowest level by: If letter of recommendation <= 4, decide "not outstanding"; otherwise, assess career plan. If career plan <= 4, decide "not outstanding"; otherwise, decide "outstanding". |
| Five cues | Like the four-cue tree but replace its lowest level by: If career plan > 4, decide "outstanding"; otherwise, assess cooperation/network. If cooperation/network <= 4, decide "not outstanding"; otherwise, decide "outstanding". |
| Six cues | Like the five-cue tree but replace its lowest level by: If cooperation/network <= 4, decide "not outstanding"; otherwise, assess clarity of project goal. If clarity of project goal <= 4, decide "not outstanding"; otherwise, decide "outstanding". |

Figure A1. Performance of fast-and-frugal trees and logistic regressions in training (238 judgments) and testing (236 judgments). Stratified random sampling was used to create the two sets (each set: 50% judgments from the first referees, 50% judgments from the second referees).

| | Training set | | | | | Testing set | | | | |
|---|---|---|---|---|---|---|---|---|---|---|
| Decision rule | #Frug | %Frug | Acc | Sens | Spec | #Frug | %Frug | Acc | Sens | Spec |
| Fast-and-frugal trees | | | | | | | | | | |
| 1 cue | 1 | 92% | .82 | .90 | .75 | 1 | 92% | .83 | .87 | .79 |
| 2 cues | 1.6 | 88% | .86 | .79 | .92 | 1.5 | 88% | .87 | .78 | .97 |
| 3 cues | 1.7 | 87% | .85 | .88 | .82 | 1.7 | 87% | .85 | .84 | .86 |
| 4 cues | 1.8 | 86% | .86 | .82 | .90 | 1.8 | 86% | .87 | .81 | .94 |
| 5 cues | 1.9 | 85% | .87 | .88 | .85 | 1.8 | 86% | .86 | .81 | .90 |
| 6 cues | 1.8 | 86% | .87 | .80 | .92 | 1.8 | 86% | .87 | .78 | .96 |
| Logistic regression | | | | | | | | | | |
| 2 cues (log-odds) | 2 | 85% | .87 | .96 | .79 | 2 | 85% | .84 | .88 | .80 |
| 2 cues (cue validity) | 2 | 85% | .87 | .81 | .92 | 2 | 85% | .87 | .80 | .95 |
| All cues | 13 | 0% | .94 | .95 | .92 | 13 | 0% | .89 | .86 | .92 |

Note: #Frug = absolute frugality; %Frug = relative frugality; Acc = accuracy; Sens = sensitivity; Spec = specificity.

*Papers)*, 1647–1661. https://doi.org/10.18653/v1/N18-1149

Langfeldt, L., & Scordato, L. (2016). *Efficiency and flexibility in research funding. A comparative study of funding instruments and review criteria*. Nordic Institute for Studies in Innovation, Research and Education.

Laskey, K., & Martignon, L. (2014). Comparing fast and frugal trees and Bayesian networks for risk assessment. In K. Makar, B. de Sousa, & R. Gould (Eds.), *Proceedings of the Ninth International Conference on Teaching Statistics*. International Statistical Institute.

Lee, C. J. (2015). Commensuration bias in peer review. *Philosophy of Science*, *82*(5), 1272–1283. https://doi.org/10.1086/683652

Lee, C. J., Sugimoto, C. R., Zhang, G., & Cronin, B. (2013). Bias in peer review. *Journal of the American Society for Information Science and Technology*, *64*(1), 2–17. https://doi.org/10.1002/asi.22784

Lindner, M. D., Vancea, A., Chen, M.-C., & Chacko, G. (2016). NIH peer review: Scored review criteria and overall impact. *American Journal of Evaluation*, *37*(2), 238–249. https://doi.org/10.1177/1098214015582049

Margherita, A., Elia, G., & Petti, C. (2022). What is quality in research? Building a framework of design, process and impact attributes and evaluation perspectives. *Sustainability*, *14*(5), 3034. https://doi.org/10.3390/su14053034

Mårtensson, P., Fors, U., Wallin, S. B., Zander, U., & Nilsson, G. H. (2016). Evaluating research: A multidisciplinary approach to assessing research practice and quality. *Research Policy*, *45*(3), 593–603. https://doi.org/10.1016/j.respol.2015.11.009

Martignon, L., & Hoffrage, U. (1999). Why does one-reason decision making work? A case study in ecological rationality. In G. Gigerenzer, P. M. Todd, & ABC Research Group (Eds.), *Simple heuristics that make us smart* (pp. 119–140). Oxford University Press.

Martignon, L., Katsikopoulos, K. V., & Woike, J. K. (2008). Categorization with limited resources: A family of simple heuristics. *Journal of Mathematical Psychology*, *52*(6), 352–361. https://doi.org/10.1016/j.jmp.2008.04.003

Martignon, L., Katsikopoulos, K. V., & Woike, J. K. (2012). Naïve, fast, and frugal trees for classification. In P. M. Todd, G. Gigerenzer, & ABC Research Group (Eds.), *Ecological rationality. Intelligence in the world* (pp. 360–378). Oxford University Press.

Martignon, L., Vitouch, O., Takezawa, M., & Forster, M. R. (2003). Naive and yet enlightened: From natural frequencies to fast and frugal decision trees. In D. Hardman & L. Macchi (Eds.), *Thinking: Psychological perspectives on reasoning, judgment, and decision making* (pp. 189–211). John Wiley and Sons.

Meehl, P. E. (1954). *Clinical versus statistical prediction: A theoretical analysis and a review of the evidence*. University of Minnesota Press.

Ochsner, M., Hug, S. E., & Daniel, H. D. (2013). Four types of research in the humanities: Setting the stage for research quality criteria in the humanities. *Research Evaluation*, *22*(2), 79–92. https://doi.org/10.1093/reseval/rvs039

Phillips, N. D., Neth, H., Woike, J. K., & Gaissmaier, W. (2017). FFTrees: A toolbox to create, visualize, and evaluate fast-and-frugal decision trees. *Judgment and Decision Making*, *12*(4), 344–368.